\documentclass[aps,preprint]{revtex4}%
\usepackage{amsfonts}%
\usepackage{amsmath}%
\usepackage{amssymb}%
\usepackage{graphicx}
\begin{document}
\preprint{HEP/123-qed}
\title[Short title for running header]{Improved Nucleon Properties in the Extended Quark Sigma Model}
\author{M. Abu-Shady}
\affiliation{Department of Mathematics, Faculty of Science, Menoufia University, Egypt}
\author{}
\affiliation{}
\author{}
\affiliation{}

\begin{abstract}
{The quark sigma model describes the quarks interacting via exchange the pions
and sigma meson fields. A new version of mesonic potential is suggested in the
frame of some aspects of the quantum chromodynamics (QCD). The field equations
have been solved in the mean-field approximation for the hedgehog baryon
state. The obtained results are compared with previous works and other models.
We conclude that the suggested mesonic potential successfully calculates
nucleon properties.}.

\end{abstract}
\volumeyear{year}
\volumenumber{number}
\issuenumber{number}
\eid{identifier}
\date[Date text]{date}
\received[Received text]{date}

\revised[Revised text]{date}

\accepted[Accepted text]{date}

\published[Published text]{date}

\startpage{101}
\endpage{102}
\maketitle
\tableofcontents

\section{$\mathbf{Introduction}$}

The description of the processes involving strong interactions is very
difficult in the frame of the quantum chromodynamics (QCD) due to its
non-abelian color and flavor structure and strong coupling constants. These
effective models, like quark sigma model, which are constructed in such a way
as to respect general properties from the more fundamental theory (QCD), such
as the chiral symmetry and its spontaneous breaking [1]. It is known that the
linear sigma model of Gell-Mann and Levy [2] does not always give the correct
phenomenology such as the value of the isoscalar pion-nucleon scattering
length is too large as in Refs. [3-5]. Birse and Banerjee [3] constructed
equations of motion treating both $\sigma$ and $\mathbf{\pi}$ fields as
time-independence classical fields and the quarks in hedgehog spinor state.
This work is reexamined by Broniowski and Banerjee [4] with corrected
numerical errors in Ref. [3]. Birse [5] generalized this mean-field
approximation to include angular momentum and isospin projection.

Recently, the mesons play an important role for improving the nucleon
properties in the chiral quark models. In the framework of the perturbative
chiral quark model [6, 7] which extended to include the kaon and eta mesons
cloud contributions to analyze the electromagnetic structure of nucleon.
Horvat et al. [8] applied Tamm-Dancoff method to the chiral quark model which
extended to include additional degrees of freedom as a pseudoscalar isoscalar
field and a triplet of scalar isovector to provide a better description of
nucleon properties. In Refs. [9-11], the authors analyzed a particular
extension of the linear sigma model coupled to valence quarks in which
contained an additional term with gradients of the chiral fields and
investigated the dynamically consequence of this term and its relevant to the
phenomenology. In addition, Rashdan et al. [12, 13] 1and Abu-shady [14]
increased the order mesonic interactions in the chiral quark sigma model using
mean-field approximation to improve nucleon properties.

The aim of the paper is to introduce the suggested mesonic potential to
improve nucleon properties and avoid the difficulty which found in the
previous works. The paper is organized as follow: In the following Section, we
review briefly the linear sigma model.\ The higher-order mesonic interactions
are studied in details in Sec. 3. The numerical calculations and the
discussion of results are presented in Secs. 4 and 5, respectively.

\section{The Chiral-Quark Sigma Model}

Brise and Banerjee [3] described the interactions of quarks via\ the exchange
of $\sigma$ and $\mathbf{\pi}$ - meson fields. The Lagrangian density is
\begin{equation}
L\left(  r\right)  =i\overline{\Psi}\partial_{\mu}\gamma^{\mu}\Psi+\frac{1}%
{2}\left(  \partial_{\mu}\sigma\partial^{\mu}\sigma+\partial_{\mu}\mathbf{\pi
}.\partial^{\mu}\mathbf{\pi}\right)  +g\overline{\Psi}\left(  \sigma
+i\gamma_{5}\mathbf{\tau}.\mathbf{\pi}\right)  \Psi-U_{1}\left(
\sigma,\mathbf{\pi}\right)  ,\tag{1}%
\end{equation}
with
\begin{equation}
U_{1}\left(  \sigma,\mathbf{\pi}\right)  =\frac{\lambda^{2}}{4}\left(
\sigma^{2}+\mathbf{\pi}^{2}-\nu^{2}\right)  ^{2}+m_{\pi}^{2}f_{\pi}%
\sigma,\tag{2}%
\end{equation}
is the meson-meson interaction potential where the $\Psi,\sigma$ and
$\mathbf{\pi}$ are the quark, sigma, and pion fields, respectively. In the
mean-field approximation, the meson fields treat as time-independent classical
fields. This means that we replace the power and the products of the meson
fields by the corresponding powers and the products of their expectation
values. In Eq. (2), the meson-meson interactions leads to the hidden chiral
symmetry $SU(2)\times SU(2)$ with $\sigma\left(  r\right)  $ taking on a
vacuum expectation value \ \ \ \ \ \ \
\begin{equation}
\ \ \ \ \ \ \left\langle \sigma\right\rangle =-f_{\pi},\tag{3}%
\end{equation}
where $f_{\pi}=92.4$ MeV is the pion decay constant. The final \ term in Eq.
(2) is included to break the chiral symmetry explicitly. It leads to the
partial conservation of axial-vector current (PCAC). The parameters
$\lambda^{2},\nu^{2}$ can be expressed in terms of$\ f_{\pi}$ and the masses
of mesons as,
\begin{equation}
\lambda^{2}=\frac{m_{\sigma}^{2}-m_{\pi}^{2}}{2f_{\pi}^{2}},\tag{4}%
\end{equation}%
\begin{equation}
\nu^{2}=f_{\pi}^{2}-\frac{m_{\pi}^{2}}{\lambda^{2}}.\tag{5}%
\end{equation}

\section{ The Chiral Higher-Order Quark Sigma Model}

The Lagrangian density of the extended linear sigma model which describes the
interactions between quarks via\ the $\sigma$ and $\mathbf{\pi}$ mesons
$\left[  14\right]  $
\begin{equation}
L\left(  r\right)  =i\overline{\Psi}\gamma_{\mu}\partial^{\mu}\Psi+\frac{1}%
{2}\left(  \partial_{\mu}\sigma\partial^{\mu}\sigma+\partial_{\mu}\mathbf{\pi
}.\partial^{\mu}\mathbf{\pi}\right)  +g\overline{\Psi}\left(  \sigma
+i\gamma_{5}\mathbf{\tau}.\mathbf{\pi}\right)  \Psi-U_{2}\left(
\sigma,\mathbf{\pi}\right)  ,\tag{6}%
\end{equation}
with%

\begin{align}
U_{2}\left(  \sigma,\mathbf{\pi}\right)   & =\frac{\lambda_{1}^{2}}{4}\left(
\sigma^{2}+\mathbf{\pi}^{2}-\nu_{1}^{2}\right)  ^{2}+\frac{\lambda_{2}^{2}}%
{4}\left(  \left(  \sigma^{2}+\mathbf{\pi}^{2}\right)  ^{2}-\nu_{2}%
^{2}\right)  ^{2}\tag{7}\\
& +m_{\pi}^{2}f_{\pi}\sigma\text{.}\nonumber
\end{align}
It is clear that potential satisfies the chiral symmetry when $m_{\pi
}\rightarrow0$. In the original model [3], the higher-order term in Eq. 7 is
excluded by the requirement of renormalizability. Since we are going to use
Eq. (7) as an approximating effective model. The model did not need and should
not be renormalizable as in Ref. [9]. By using the PCAC and the minimization
conditions of mesonic potential $\left[  14\right]  $, we obtain
\begin{equation}
\lambda_{1}^{2}=\frac{m_{\sigma}^{2}-m_{\pi}^{2}}{4f_{\pi}^{2}}%
,\ \ \ \ \ \ \ \ \nu_{1}^{2}=f_{\pi}^{2}-\frac{m_{\pi}^{2}}{\lambda_{1}^{2}%
},\tag{8}%
\end{equation}%
\begin{equation}
\lambda_{2}^{2}=\frac{m_{\sigma}^{2}-3m_{\pi}^{2}}{16f_{\pi}^{6}}%
,\ \ \ \nu_{2}^{2}=f_{\pi}^{4}-\frac{m_{\pi}^{2}}{2\lambda_{2}^{2}f_{\pi}^{2}%
}.\tag{9}%
\end{equation}
Now we can expand the extremum with the shifted field defined as
\begin{equation}
\sigma=\sigma^{\prime}-f_{\pi},\tag{10}%
\end{equation}
substituting Eq. (10) into Eq. (6), we get%

\begin{align}
L\left(  r\right)   & =i\overline{\Psi}\gamma_{\mu}\partial^{\mu}\Psi+\frac
{1}{2}\left(  \partial_{\mu}\sigma^{\prime}\partial^{\mu}\sigma^{\prime
}+\partial_{\mu}\mathbf{\pi}.\partial^{\mu}\mathbf{\pi}\right)  -g\overline
{\Psi}f_{\pi}\Psi+g\overline{\Psi}\sigma^{\prime}\Psi+ig\overline{\Psi
}\mathbf{\gamma}_{5}.\mathbf{\pi}\Psi\nonumber\\
& -U_{2}\left(  \sigma^{\prime},\mathbf{\pi}\right)  ,\tag{11}%
\end{align}
with
\begin{align}
U_{2}\left(  \sigma^{\prime},\mathbf{\pi}\right)   & =\frac{\lambda_{1}^{2}%
}{4}(\left(  \sigma^{\prime}-f_{\pi})^{2}+\mathbf{\pi}^{2}-\nu_{1}^{2}\right)
^{2}+\frac{\lambda_{2}^{2}}{4}\left(  \left(  (\sigma^{\prime}-f_{\pi}%
)^{2}+\mathbf{\pi}^{2}\right)  ^{2}-\nu_{2}^{2}\right)  ^{2}\nonumber\\
& +m_{\pi}^{2}f_{\pi}(\sigma^{\prime}-f_{\pi}).\tag{12}%
\end{align}
The time-independent fields $\sigma^{^{\prime}}\left(  r\right)  \,\,$and
$\mathbf{\pi}\left(  r\right)  $ satisfy the Euler$-$Lagrange equations, and
the quark wave function satisfies the Dirac eigenvalue equation. Substituting
Eq. (11) in Euler$-$Lagrange equation, we get
\begin{align}
\square\sigma^{\prime}  & =g\overline{\Psi}\Psi-\lambda_{1}^{2}(f_{\pi}%
-\sigma^{\prime})((\sigma^{\prime}-f_{\pi})^{2}+\mathbf{\pi}^{2}-\nu_{1}%
^{2})-\nonumber\\
& 2\lambda_{2}^{2}(f_{\pi}-\sigma^{\prime})\left(  (\sigma^{\prime}-f_{\pi
})^{2}+\mathbf{\pi}^{2}\right)  (\left(  (\sigma^{\prime}-f_{\pi}%
)^{2}+\mathbf{\pi}^{2}\right)  ^{2}-\nu_{2}^{2})-m_{\pi}^{2}f_{\pi},\tag{13}%
\end{align}%
\begin{align}
\square\mathbf{\pi}  & =ig\overline{\Psi}\gamma_{5\cdot}\mathbf{\tau}%
\Psi-\lambda_{1}^{2}((\sigma^{\prime}-f_{\pi})^{2}+\mathbf{\pi}^{2}-\nu
_{1}^{2}))\mathbf{\pi}-\nonumber\\
& 2\lambda_{2}^{2}\mathbf{\pi}\left(  (\sigma^{\prime}-f_{\pi})^{2}%
+\mathbf{\pi}^{2}\right)  (\left(  (\sigma^{\prime}-f_{\pi})^{2}+\mathbf{\pi
}^{2}\right)  ^{2}-\nu_{2}^{2}),\tag{14}%
\end{align}
where $\mathbf{\tau}$ refers to Pauli isospin matrices, $\gamma_{5}=\left(
\begin{array}
[c]{cc}%
0 & 1\\
1 & 0
\end{array}
\right)  $. Including the color degree of freedom, one has $g\overline{\Psi
}\Psi\rightarrow N_{c}g\overline{\Psi}\Psi$ where $N_{c}=3$ colors. Thus
\begin{equation}
\Psi\left(  r\right)  =\frac{1}{\sqrt{4\pi}}\left[
\begin{array}
[c]{c}%
u\left(  r\right) \\
iw\left(  r\right)
\end{array}
\right]  \qquad\text{and}\qquad\bar{\Psi}\left(  r\right)  =\frac{1}%
{\sqrt{4\pi}}\left[
\begin{array}
[c]{cc}%
u\left(  r\right)  & iw\left(  r\right)
\end{array}
\right]  ,\tag{15}%
\end{equation}
then
\begin{align}
\rho_{s}  & =N_{c}\overline{\Psi}\Psi=\frac{3g}{4\pi}\left(  u^{2}%
-w^{2}\right)  ,\tag{16}\\
\rho_{p}  & =iN_{c}\overline{\Psi}\gamma_{5}\mathbf{\tau}\Psi=\frac{3g}{2\pi
}(uw),\tag{17}\\
\rho_{v}  & =\frac{3g}{4\pi}\left(  u^{2}+w^{2}\right)  ,\tag{18}%
\end{align}
where $\rho_{s},$ $\rho_{p}$ and $\rho_{v}$ are sigma, pion and vector
densities, respectively. These equations are subject to the boundary
conditions as follows,
\begin{equation}
\sigma\left(  r\right)  {\sim}-f_{\pi},\ \ \ \ \pi\left(  r\right)  {\sim
}0\text{ \ \ \ \ at }r\rightarrow\infty\text{.}\tag{19}%
\end{equation}
By using hedgehog ansatz [12], where
\begin{equation}
\mathbf{\pi}\left(  r\right)  =\pi\left(  r\right)  \overset{\symbol{94}%
}{\mathbf{r}}.\tag{20}%
\end{equation}
The chiral Dirac equation for the quarks is [12]
\begin{equation}
\frac{du}{dr}=-P\left(  r\right)  u+\left(  W+m_{q}-S(r)\right)  w,\tag{21}%
\end{equation}
where the scalar potential $S(r)=g\left\langle \sigma^{\prime}\right\rangle $,
the pseudoscalar potential $P(r)=\left\langle \mathbf{\pi}\cdot\mathbf{\hat
{r}}\right\rangle $, and $W$ is the eigenvalue of the quarks spinor $\Psi$%
\begin{equation}
\frac{dw}{dr}=-\left(  W-m_{q}+S(r)\right)  u-\left(  \frac{2}{r}-P\left(
r\right)  \right)  w.\tag{22}%
\end{equation}

\section{Numerical Calculations and Discussions}

\subsection{The scalar field $\sigma^{\prime}$}

To solve Eq. (13), we integrate a suitable Green's function over the source
fields as in Refs. $\left[  12,13\right]  .$\ Thus
\begin{align}
\sigma^{\prime}\left(  \mathbf{r}\right)   & =\int d^{3}\mathbf{r}^{\prime
}D_{\sigma}(\mathbf{r-\grave{r}})[g\rho_{s}(\mathbf{\grave{r}})-\lambda
_{1}^{2}(f_{\pi}-\sigma^{\prime})((\sigma^{\prime}-f_{\pi})^{2}+\mathbf{\pi
}^{2}-\nu_{1}^{2})-\nonumber\\
& 2\lambda_{2}^{2}(f_{\pi}-\sigma^{\prime})\left(  (\sigma^{\prime}-f_{\pi
})^{2}+\mathbf{\pi}^{2}\right)  (\left(  (\sigma^{\prime}-f_{\pi}%
)^{2}+\mathbf{\pi}^{2}\right)  ^{2}-\nu_{2}^{2})-m_{\pi}^{2}f_{\pi
}],\;\;\;\;\;\;\;\;\;\;\;\;\ \tag{23}%
\end{align}
where
\[
D_{\sigma}(\mathbf{r-\grave{r}})=\frac{1}{4\pi\left\vert \mathbf{r-\grave{r}%
}\right\vert }\exp(-m_{\sigma}\left\vert \mathbf{r-\grave{r}}\right\vert ),\;
\]
the scalar field is spherical in this model so we only need the $l=0$\ term
\begin{equation}
D_{\sigma}\left(  \mathbf{r-\grave{r}}\right)  =\frac{1}{4\pi}\sinh\left(
m_{\sigma}r_{<}\right)  \frac{\exp\left(  -m_{\sigma}r_{>}\right)  }{r_{>}%
},\;\;\tag{24}%
\end{equation}
therefore
\begin{align}
\sigma^{\prime}\left(  \mathbf{r}\right)   & =m_{\sigma}\int\limits_{0}%
^{\infty}r^{\prime2}dr^{\prime}(\frac{\sinh\left(  m_{\sigma}r_{>}\right)
\exp\left(  -m_{\sigma}r_{>}\right)  }{m_{\sigma}r_{>}})[g\rho_{s}%
(\mathbf{\grave{r}})-\tag{25}\\
& \lambda_{1}^{2}(f_{\pi}-\sigma^{\prime})((\sigma^{\prime}-f_{\pi}%
)^{2}+\mathbf{\pi}^{2}-\nu_{1}^{2})-2\lambda_{2}^{2}(f_{\pi}-\sigma^{\prime
})\left(  (\sigma^{\prime}-f_{\pi})^{2}+\mathbf{\pi}^{2}\right)
\times\nonumber\\
& \times(\left(  (\sigma^{\prime}-f_{\pi})^{2}+\mathbf{\pi}^{2}\right)
^{2}-\nu_{2}^{2})-m_{\pi}^{2}f_{\pi}]\text{.}\nonumber
\end{align}
Note that this form is implicit in the solution of $\sigma^{\prime}$involves
integrals over the unknown $\sigma^{\prime}$ itself. We will solve this
implicit integral equation by iterating to self-consistency.

\subsection{The pion field $\mathbf{\pi}$}

To solve Eq. (14), we integrate a suitable Green's function over the source
fields. We use the $l=1$\ component of the pion Green's function. Thus
\begin{align}
\;\;\;\ \ \mathbf{\pi}\left(  r\right)   & =m_{\pi}\int_{0}^{\infty}%
r^{\prime2}dr^{\prime}\frac{[-\sinh\left(  m_{\pi}r_{<}\right)  +m_{\pi}%
r_{<}\cosh\left(  m_{\pi}r_{<}\right)  ]}{\left(  m_{\pi}r_{>}\right)  ^{2}%
}\times
\;\;\;\;\;\;\;\;\;\;\;\;\;\;\;\;\;\;\;\;\;\;\;\;\;\;\;\;\;\;\;\;\;\tag{26}\\
\;\;\;\;\;\;  & [(1+\frac{1}{m_{\pi}r_{>}})\frac{\exp\left(  -m_{\pi}%
r_{>}\right)  }{m_{\pi}r_{>}})(g\rho_{p}-\lambda_{1}^{2}((\sigma^{\prime
}-f_{\pi})^{2}+\mathbf{\pi}^{2}-\nu_{1}^{2}))\mathbf{\pi-}\nonumber\\
& 2\lambda_{2}^{2}\mathbf{\pi}\left(  (\sigma^{\prime}-f_{\pi})^{2}%
+\mathbf{\pi}^{2}\right)  (\left(  (\sigma^{\prime}-f_{\pi})^{2}+\mathbf{\pi
}^{2}\right)  ^{2}-\nu_{2}^{2})].\nonumber
\end{align}
We have solved Dirac Eqs. (21), (22) using fourth-order Rung Kutta method. Due
to the implicit nonlinearly of these Eqs. (13), (14) it is necessary to
iterate the solution until self-consistency is achieved. To start this
iteration process, we could use the chiral circle form for the meson fields
[12, 13]:
\begin{equation}
S(r)=m_{q}(1-\cos\theta),\text{ }P(r)=-m_{q}\sin\theta,\tag{27}%
\end{equation}
where $\theta=\tanh r$.\newpage

\ \ \ \ \ \ \ \ \ 

\subsection{ \ The Properties of the Nucleon}

The proton and neutron magnetic moments are given by [3]
\begin{equation}
\mu_{p,n}=<P\uparrow\left\vert \int\frac{1}{2}\mathbf{r}\times\mathbf{j}%
_{\varepsilon M}(\mathbf{r})d^{3}\mathbf{r}\right\vert P\uparrow>,\tag{28}%
\end{equation}

where, the electromagnetic current is
\begin{equation}
j_{\epsilon M}(\mathbf{r})=\bar{\Psi}\left(  \mathbf{r}\right)  \mathbf{\gamma
}\left(  \frac{1}{6}+\frac{\tau_{3}}{2}\right)  \Psi(\mathbf{r})-\varepsilon
_{\alpha\beta_{3}}\pi_{\alpha}\left(  \mathbf{r}\right)  \mathbf{\nabla}%
\pi_{\beta}\left(  \mathbf{r}\right)  ,\tag{29}%
\end{equation}

such that
\begin{equation}
\left(  \mathbf{j}_{\epsilon M}(\mathbf{r})\right)  _{nucleon}=\bar{\Psi
}\left(  \mathbf{r}\right)  \mathbf{\gamma}\left(  \frac{1}{6}+\frac{\tau_{3}%
}{2}\right)  \Psi\left(  \mathbf{r}\right)  ,\tag{30}%
\end{equation}%
\begin{equation}
\left(  \mathbf{j}_{\epsilon M}(\mathbf{r})\right)  _{meson}=-\epsilon
_{\alpha\beta3}\pi_{\alpha}\left(  \mathbf{r}\right)  \mathbf{\nabla}%
\pi_{\beta}\left(  \mathbf{r}\right)  .\tag{31}%
\end{equation}
The nucleon axial-vector coupling constant is found from%
\begin{equation}
\frac{1}{2}g_{A}(0)=\left\langle P\uparrow\left\vert \int d^{3}rA_{3}%
^{z}(\mathbf{r})\right\vert P\uparrow\right\rangle ,\tag{32}%
\end{equation}
where the z-component of the axial vector current is given by
\begin{equation}
A_{3}^{z}(\mathbf{r})=\bar{\Psi}\left(  \mathbf{r}\right)  \frac{1}{2}%
\gamma_{5}\gamma^{3}\tau_{3}\Psi\left(  \mathbf{r}\right)  -\sigma\left(
\mathbf{r}\right)  \frac{\partial}{\partial z}\pi_{3}\left(  \mathbf{r}%
\right)  +\pi_{3}\left(  \mathbf{r}\right)  \frac{\partial}{\partial z}%
\sigma\left(  \mathbf{r}\right)  .\tag{33}%
\end{equation}
The pion-nucleus $\sigma$ commutator is defined
\begin{equation}
\sigma(\pi N)=\left\langle P\uparrow\left\vert \int\sigma^{\prime}%
(\mathbf{r})d^{3}r\right\vert P\uparrow\right\rangle .\tag{34}%
\end{equation}
In calculation of $\sigma(\pi N),$ we replace $\sigma^{\prime}(\mathbf{r}) $
by $\frac{j_{\sigma}(\mathbf{r})}{m_{\sigma}^{2}}$ where $j_{\sigma
}(\mathbf{r})$ is the source current defined by%

\[
(\square+m_{\sigma}^{2})\sigma^{\prime}=j_{\sigma}(\mathbf{r}).
\]
\quad\ \ The hedgehog mass is calculated in details in Refs. $\left[
12,13\right]  $.\ \ \ \ \ \ \ \ \ \ \ \ \ \ \ \ \ \ \ \ \ \ \ \ \ \ \ \ \ \ \ \ \ \ \ \ \ \ \ \ \ \ \ \ \ \ \ \ \ \ \ \ \ \ \ \ \ \ \ \ \ \ \ \ \ \ \ \ \ \ \ \ \ \ \ \ \ \ \ \ \ \ \ \ \ \ \ \ \ \ \ \ \ \ \ \ \ \ \ \ \ \ \ \ \ \ \ \ \ \ \ \ \ \ \ \ \ \ \ \ \ 

\subsection{Discussion of the Results}

The set of equations (13-22) are numerically solved by the iteration method as
Refs. [12-14] for different values of the sigma and quark masses. The
dependence of the nucleon properties on the sigma and the quark masses are
listed in the tables (1), (2), (3), and (4). In Table (1), we note that the
hedgehog mass, the magnetic moments of the proton and neutron, and the sigma
commutator increase by increasing sigma mass. We obtain a good value of the
hedgehog mass equals to 1090 MeV which closed to experimental data 1086 MeV.
In Table (2), we examine the effect of quark mass on the nucleon properties.
We note\ that the hedgehog mass decreases with increasing quark mass. This
interpreted that an increase in the quark mass leads to increase in the
coupling constant $(g=\frac{m_{q}}{f_{\pi}})$. Therefore, the coupling between
meson and the quark more tight, leading the decrease in the hedgehog mass as
in Refs. [3, 12, 13]. Also, we note that the magnetic moments of proton and
neutron increase by increasing quark mass. A similar effect occurred respect
to sigma commutator $\sigma(\pi N).$ In comparison between the results in the
tables 1 and 2. We note that quark mass is more affected on nucleon properties
that the strong change of sigma mass leads to the change of nucleon properties
as in the table 1. In Table (3), we compare between the original quark model
and the higher-order quark model. We fixed all parameters in the two models to
show the effect of the higher-order mesonic interactions on the nucleon
properties. We note that the dynamic of kinetic energy of quark increases by
increasing mesonic contributions in the original quark model. In addition, the
meson-quark interaction energy decreases by increasing higher-order
interactions. We note that meson-meson interaction decreases by increasing
mesonic contributions in the original sigma model. We obtain the excellent
value of hedgehog mass $M_{H}$ $\cong1090$ MeV while we obtain $M_{H}$
$\cong1068$ MeV in the original sigma model at the same free parameters.
Therefore, an increase of the mesonic interactions improved the hedgehog mass
which closed to experimental data ($M_{H}$ $\cong1086$ MeV). The magnetic
moments of proton and neutron are improved in comparison with the original
model. Sigma commutator $\sigma(\pi N)$ is one of problems in the original
sigma model that is a largest value in comparison with data. By increasing
mesonic contributions in the original sigma model. This value reduced from 126
MeV to 78 MeV. Therefore, the value improved about 38 \% and it is acceptable
agreement with experimental data. The quantity $g_{A}(0)$ is improve in
comparsion with the original model but still a large value in comparing with
experimental data $\left(  1.25\right)  $. Since the $g_{A}(0)$ depends on the
meson fields only not on the coupling of higher-order term in the extended
sigma model. Therefore, we need to add a vector meson to our model to improve
this quantity, which will be a future paper.

\textbf{Table (1).} Values of magnetic moments of proton and neutron, the
hedgehog mass{\small \ }$M_{B}${\small , } and $\sigma(\pi N)$ for $m_{\pi
}=139.6$ MeV$,m_{q}=500$ MeV$,$ $f_{\pi}=92.4\,$MeV$.$ All quantities in MeV.%

\begin{tabular}
[c]{|l|l|l|l|l|}\hline
$m_{\sigma}\left(  \text{MeV}\right)  $ & 600 & 700 & 800 & 900\\\hline
Hedgehog mass $M_{B}$ & 1090.92 & 1108.98 & 1125.54 & 1139.27\\\hline
Total moment proton $\mu_{p}\left(  N\right)  $ & 2.8456 & 2.8641 & 2.8643 &
2.8646\\\hline
Total moment neutron $\mu_{n}\left(  N\right)  $ & -2.2076 & -2.2374 &
-2.2494 & -2.259\\\hline
$\sigma(\pi N)$ & 77.025 & 78.158 & 78.440 & 78.770\\\hline
\end{tabular}

\bigskip

\textbf{Table (2).} Values of magnetic moments of proton and neutron, the
hedgehog mass{\small \ }$M_{B}${\small , } and $\sigma(\pi N)$ for $m_{\pi
}=139.6$ MeV$,m_{\sigma}=600$ MeV$,$ $f_{\pi}=92.4\,$MeV$.$ All quantities in MeV.%

\begin{tabular}
[c]{|l|l|l|l|l|l|l|}\hline
$m_{q}\left(  \text{MeV}\right)  $ & 400 & 420 & 440 & 460 & 480 & 500\\\hline
Hedgehog mass $M_{B}$ & 1230 & 1210 & 1185 & 1157 & 1124 & 1089\\\hline
Total moment proton $\mu_{p}\left(  N\right)  $ & 2.574 & 2.653 & 2.719 &
2.775 & 2.823 & 2.845\\\hline
Total moment neutron $\mu_{n}\left(  N\right)  $ & -1.899 & -1.985 & -2.05 &
-2.121 & -2.175 & -2.207\\\hline
$\sigma(\pi N)$ & 49.19 & 57.57 & 64.28 & 69.79 & 74.32 & 77.02\\\hline
\end{tabular}
\bigskip

\textbf{Table(3}). Details of energy calculations of the hedgehog mass, the
magnetic moments of proton and neutron, and the sigma commutator $\sigma(\pi
N)$ for $m_{q}=500$ MeV$,m_{\pi}=139.6$ MeV$,m_{\sigma}=600$ MeV$,$ and
$f_{\pi}=92.4\,$MeV$.$ All quantities in MeV.%

\begin{tabular}
[c]{|l|l|l|}\hline
Quantity & Original Sigma Model & Higher-order Sigma Model\\\hline
Quark kinetic energy & 1166.38 & 1171.068\\\hline
Sigma kinetic energy & 353.15 & 375.038\\\hline
Pion kinetic energy & 461.85 & 451.827\\\hline
Sigma interaction energy & -165.84 & -165.975\\\hline
Pion interaction energy & -860.87 & -854.098\\\hline
Meson interaction energy & 114.0 & 113.069\\\hline
Hedgehog mass baryon & 1068.67 & 1090.92\\\hline
Total moment of proton $\mu_{p}$ & 2.89 & 2.84\\\hline
Total moment of neutron $\mu_{n}$ & -2.24 & 2.20\\\hline
$g_{A}(0)$ & 1.80 & 1.78\\\hline
$\sigma(\pi N)$ & 126.99 & 77\\\hline
\end{tabular}

\section{Comparison with Other Models}

It is interesting to compare the nucleon properties in the present approach
with the previous works and other models. The higher-order mesonic potential
was suggested in Refs. [12-14]. In Ref. [12], the sigma commutator $\sigma(\pi
N)$ is not calculated in this work. It is an essential property of nucleon
properties. In addition, the mesonic potential has a weakness point at
$m_{\pi}=0$ so the model did not satisfy the chiral limit case. We note that
the hedgehog mass improved in comparison with result of Ref. [12 ]. In Ref.
[13], the authors suggested another form of mesonic potential to avoid the
difficulty which came from $m_{\pi}=0.$ We have two advantages in comparison
with Ref. [13]. The first, our results in the present work are improved, in
particular the hedgehog mass and the $\sigma(\pi N)$. The second, the mesonic
potential in Eq. 7, has the similar form when the coupling constant of
higher-order $\lambda_{2}^{2}$ is vanished as in Eq. 2. This advantage is not
found in Ref. [13]. In Ref. [14], the author studied the effect of large pion
masses on the magnetic moments of proton and neutron only.

It is important to compare present model with other models such as the
perturbative chiral quark Model [6, 7] and the extended Skyrme model [15]. The
perturbative chiral quark model is an effective model of baryons based on
chiral symmetry. The baryon is described as a state of three localized
relativistic quarks supplemented by a pseudoscalar meson cloud as dictated by
chiral symmetry requirements. In this model, the effect of the meson cloud is
evaluated perturbatively in a systematic fashion. The model has been
successfully applied to the nucleon properties (see Table 4). We obtain
reasonable results in comparison with this model for the $\sigma(\pi N)$ which
backs to perturbative chiral quark model based on non-linear $\sigma-$ model
Lagrangian. In particular, nucleon magnetic moments are improved in comparison
with this model. Moreover, Hedgehog mass $M_{B}$ is not calculated in this
model. The original Skyrme model [16] consists of the non-linear sigma term
and the fourth-order derivative term, which guarantees the stabilization of
the soliton so that the degree of freedom of the sigma field may be replaced
by a variable chiral radius, which becomes the new dynamical degree of freedom
and plays an important role in the modified Skyrmion Lagrangian density [15],
leading to a better description of nucleon properties. In comparison with the
extended Skyrme model [15], the results obtained for the hedgehog mass have
been improved and the other properties are in agreement with this model (see
Table 4).

\textbf{Table (4). }Values of the observables calculated from the extended
linear sigma model [12, 13], the perturbative chiral quark model [6, 7], and
the extended Skyrme model [15] in comparison with the present work.%

\begin{tabular}
[c]{|l|l|l|l|l|l|l|}\hline
Quantity & Present work & \ [ 13\ ] & [6, 7] & \ [ 12\ ] & \ [15] &
Expt.\\\hline
Hedgehog mass $M_{B}$ & 1090 & 1200 & - & 1081 & 1157 & 1086\\\hline
$\mu_{p}\left(  N\right)  $ & 2.84 & 2.76 & 2.62$\pm0.02$ & 2.768 & 2.77 &
2.79\\\hline
$\mu_{n}\left(  N\right)  $ & -2.20 & -1.91 & -2.02$\pm0.02$ & -1.909 &
-2.11 & -1.91\\\hline
$\sigma(\pi N)$ & 77 & 88 & 54.7 & - & 70 & 50$\pm20$\\\hline
\end{tabular}

\section{ Conclusion}

The present calculations have shown the importance of mesonic corrections of
higher-order than that normally used in most soliton models. The obtained
results are improved in comparison with previous calculations. In addition, we
avoid the difficulty that found in the previous works. The advantage\ of the
present work that hedgehog mass is corrected and closed with data. The
magnetic moments of proton and neutron and sigma commutator $\sigma(\pi N)$
are improved in comparison with other models.

\section{\textbf{References}}

\begin{enumerate}
\item S. Gasiorowicz and D. A. Geffen, Rev. Mod. Phys. \textbf{41}, 531 (1969).

\item M. Gell-Mann, M. Levy, Nuovo Cimento \textbf{16,} 705 (1960).

\item M. Birse and M. Banerjee, Phys. Rev. D \textbf{31}, 118 (1985).

\item W. Broniowski and M. K. Banerjee, Phys. Lett. B \textbf{158}, 335 (1985).

\item M. Birse, Phys. Rev. D \textbf{33}, 1934 (1986).

\item V.~E.~Lyubovitskij, T.~Gutsche and A.~Faessler, Phys.\ Rev.\ C
\textbf{64}, 065203 (2001).

\item T. Inoue, V.~E.~Lyubovitskij, T.~Gutsche, A.~Faessler, Phys. Rev. C
\textbf{69}, 035207 (2004).

\item D. Horvat, D. Horvatic, B. Podobnik and D. Tadic, FIZIKA\ B \textbf{9,}
181 (2000).

\item W. Broniowski and B. Golli, Nucl. Phys. A \textbf{714,} 575 (2003).

\item M. Abu-Shady, Acta Phys. Polo. B \textbf{40}, 8 (2009).

\item M. Abu-Shady, Int. J. Theor. Phys. \textbf{48}, 1110 (2009).

\item M. Rashdan, M. Abu-Shady, and T.S.T Ali, Inter. J. Mod. Phys. A
\textbf{22}, 2673 (2007)

\item M. Rashdan, M. Abu-shady, and T.S.T. Ali, Int. J. Mod. Phys. E
\textbf{15, }143 ( 2006).

\item M. Abu-Shady, Phys. Atom. Nucl. \textbf{73}, 978 (2010).

\item F. L. Braghin and I. P. Cavalcante, Phys. Rev. C \textbf{67}, 065207 (2003).

\item T. H. R. Skyrme, Proc. R. Soc. London, Ser. A \textbf{260}, 127 (1961).
\end{enumerate}

\end{document}